# A dose-effect network meta-analysis model with application in antidepressants using restricted cubic splines

Tasnim Hamza[1,2], Toshi A. Furukawa[3], Nicola Orsini[4], Andrea Cipriani[5], Cynthia P Iglesias[6], Georgia Salanti[1].

[1]Institute of Social and Preventive Medicine, University of Bern, Bern, Switzerland.
[2]Graduate School for Health Sciences, University of Bern, Switzerland
[3]Department of Health Promotion and Human Behavior, and Department of Clinical Epidemiology, Graduate School of Medicine/School of Public Health, Kyoto University, Kyoto, Japan. [4]Department of Global Public Health, Karolinska Institutet, Stockholm, Sweden. [5]Department of Psychiatry, University of Oxford. [6]Department of Health Sciences, University of York, York.

**Acknowledgements:**

TH, CI, and GS are funded by the European Union's Horizon 2020 research and innovation programme under grant agreement No 825162. AC is supported by the National Institute for Health Research (NIHR) Oxford Cognitive Health Clinical Research Facility, by an NIHR Research Professorship (grant RP-2017-08-ST2-006), by the NIHR Oxford and Thames Valley Applied Research Collaboration and by the NIHR Oxford Health Biomedical Research Centre (grant BRC-1215-20005). The views expressed are those of the authors and not necessarily those of the UK National Health Service, the NIHR, or the UK Department of Health. TAF reports grants and personal fees from Mitsubishi-Tanabe, personal fees from MSD, personal fees from SONY, grants and personal fees from Shionogi, outside the submitted work; In addition, TAF has a patent 2020-548587concerning smartphone CBT apps pending, and intellectual properties for Kokoro-app licensed to Mitsubishi-Tanabe.

**Abstract**

Network meta-analysis (NMA) has been used to answer a range of clinical questions about the preferable intervention for a given condition. Although the effectiveness and safety of pharmacological agents depend on the dose administered, NMA applications typically ignore the role that drugs dosage plays in the results. This leads to more heterogeneity in the network. In this paper, we present a suite of NMA models that incorporate the dose-effect relationship (DE-NMA) using restricted cubic splines (RCS). We extend existing models into a dose-effect network meta-regression to account for study-level covariates and for groups of agents in a class-effect DE-NMA model. We apply our models to a network of aggregate data about the efficacy of 21 antidepressants and placebo for depression. We find that all antidepressants are more efficacious than placebo after a certain dose. Also, we identify the dose level at which each antidepressant's effect exceeds that of placebo and estimate the dose beyond which the effect of antidepressants no longer increases. When covariates were introduced to the model, we find that studies with small sample size tend to exaggerate antidepressants efficacy for several of the drugs. Our DE-NMA model with RCS provides a flexible approach to modelling the dose-effect relationship in multiple interventions. Decision-makers can use our model to inform treatment choice.

# 1 Introduction

Network meta-analysis (NMA) is a technique commonly used to simultaneously compare multiple agents [1–3]. Although comparison between pharmacological agents is important, in practice clinicians always prescribe drugs at a particular dose, informed by the market authorisation, licensing of the product, dose-effect studies, and their experience. It is therefore important to know not only which pharmacological agents are preferable but how their advantage depends on the dose.

Health technology assessment agencies make recommendations that should, and sometimes do, specify the recommended dose range for several competing pharmacological agents. However, without a unified methodological approach to infer the relative effects of agent-dose combinations, contradictory information might be made available. For example, the guidelines produced by the National Institute of Health and Care Excellence state that no dose dependency has been established within the therapeutic range of selective serotonin-reuptake inhibitors (SSRI) when treating people diagnosed with major depression, whereas the American Psychiatric Association guideline recommends titration up to the maximum tolerated dose.

In NMA, the first and often most challenging step is the definition of the nodes in the network with respect to the combination of agents and dose. When pharmacological agents are compared, an important decision faced early on is whether the dose of each agent is of interest, and consequently, whether the definition of each node involves the dose of the agent or not.

There are three main options when it comes to dealing with the dose of pharmacological interventions in NMA. Frequently, information about the dose is ignored and focus is placed only on the relative effects between agents (e.g. Cipriani et al. [4]). This approach may result in a network with increased heterogeneity and inconsistency. At the other end of the spectrum, one can consider each agent-dose combination as a different treatment that defines a different node in the network [5]. This detailed and larger network will inevitably be at best sparse or even disconnected. A compromise is to model the dose-effect relationship for each agent by extending the dose-effect meta-analysis models [6–8].

The dose-effect relationship expresses the change in effect over different doses. In pairwise meta-analysis, the dose-effect curves are synthesized across studies. Such analyses can be conducted using the two-stage or one-stage methods in a frequentist [9, 10] or Bayesian setting [11]. In NMA, the linear dose-effect model has been implemented [6] which, however, poorly reflects the natural dose-effect dynamics [12]. Del Giovane et al. [7] addressed that by either considering an exchangeable effect for the different doses of a certain agent or assuming the dose-effect relationship as a monotonic, linear, or random walk. More recently, Mawdesly et al. [8] have extended NMA to incorporate the Emax dose-effect model which is commonly used in pharmacometrics when determining the optimal dose. In clinical practice and for decision making, more flexibility in the assumed dose-effect shapes is desirable to better reflect a range of possible biological mechanisms of the various pharmacological agents.

With this paper, we aim to contribute to the growing literature about dose-effect models by describing a generic and flexible dose-effect NMA (DE-NMA) model with restricted cubic splines (RCS). Recent simulations showed that the RCS successfully capture a large range of functional shapes [11]. As residual heterogeneity and inconsistency (beyond what can be explained by different dosages) can occur, we extend the model into a dose-effect network meta-regression by incorporating study-level covariates.

The article is structured as follows. First, we present our motivating example. In Section 3, we present the DE-NMA model and two extensions: the DE-NMR and a DE-NMA that includes class effects. Next, we apply the models to the antidepressants network, and we then present the results. In Section 4, we discuss the strengths and limitations of the models, and we discuss other methods to estimate the dose-response shape, such as fractional polynomials.

The analysis code is implemented using Just Another Gibbs Sampler (JAGS) [13] and R [14], and it is made available at Zenodo [15].

## 2 Example: Comparing the efficacy of 21 antidepressants

We illustrate our different models using a network of double-blind fixed-dose randomised controlled trials (RCT) that compare antidepressants for depression (see Figure 1a and Appendix Figure 1). The primary outcome is efficacy measured as the total number of patients who had more than 50% reduction in symptoms (response rate) [16]. The participants of the included studies were adults diagnosed with unipolar major depressive disorder. The dataset is a superset of one used to compare 21 antidepressants and placebo according to their efficacy, acceptability, and safety [4]. In that NMA, Cipriani et al. synthesized only arms with agents administered at approved doses (as fixed or flexible schedule), while we included all trial arms regardless of the dosage. More details about inclusion criteria, the search strategy, data extraction, and risk of bias in these studies can be found in Cipriani et al. [4].

Our dataset includes 170 RCTs comparing 21 antidepressants with placebo or another active treatment. Trazodone is excluded from the primary analysis (because only one dose level is examined in the included studies), yet included in the class-effect model (it belongs with nefazodone in the same class, serotonin antagonist and reuptake inhibitor [SARIs]). The trials report 457 different fixed dose-per-drug treatments and include 54,048 participants. In Appendix Table 1, we summarise the number of events, the sample size, the number of studies, the number of different doses, and the class for each drug. We present the distribution of observed doses per drug in Figure 1b.

A subset of our data (only SSRIs except fluvoxamine) have been previously analysed using pairwise dose-effect meta-analysis, thus ignoring the differences between the individual drugs (using frequentist [17] or Bayesian dose-effect meta-analysis [11]).

# 3 Methods

We first present the DE-NMA model, and then we extend it by incorporating covariates or by assuming class effects between the exposures. As most studies in any NMA are RCTs, we assume the case where each arm of a trial has been randomized to an agent at a different dose. We also present the model assuming a dichotomous outcome using RCS for the association between dose and effects. However, the model can be easily adapted for any assumed shape (e.g., linear, quadratic, etc.) and any type of outcome.

## 3.1 Notation

Table 1 summarizes the notation we used. Suppose we compared $K$ agents ($k = 1,..,K$) in $ns$ studies ($i = 1,..,ns$) that report the dose level $j$. For each dose, $x_{ijk}$, we observe the number of events $r_{ijk}$ and the sample size $n_{ijk}$ (dichotomous outcome). Additionally, we have information on a study-level covariate $Z_i$. In the class effect model, the $c$ index refers to the class of the agent. Note that we differentiated between agent and treatment when the latter refers to the given dose of a certain agent.

## 3.2 Dose-effect network meta-analysis model with placebo arm

We defined the DE-NMA model as an extension of the standard NMA. We describe it as a hierarchical model with three layers; we first estimated the dose-effect association within each study, and then we synthesised the shapes across studies and across agents. For simplicity, we present the model assuming that the network includes placebo.

### *3.2.1 Dose-effect model within each study*

For ease of understanding and notation, we begin our description of a dose-effect model for a network of trials that have "placebo" as a common comparator. We relax this initial assumption at the end of Section 3.2.1, when we describe a dose-effect model for a network of studies with different controls.

Let us assume that within each study $i$, the number of events follows a binomial distribution

$$r_{ijk} \sim \text{Binomial}(p_{ijk}, n_{ijk})$$

with $p_{ijk}$ being the probability of an event to occur. We choose a transformation of these probabilities based on the measure of the relative effect that we are interested in. We set the transformation to the logit transformation for odds ratio (OR)

$$\text{logit}(p_{ijk}) = \begin{cases} u_i, & \text{placebo} \\ u_i + \delta_{ijk}, & \text{active agent} \end{cases}$$

$u_i$ is the log-odds of the event on the placebo arm in study $i$. The term $\delta_{ijk}$ denotes the underlying parameter for the effect of agent $k$ in study $i$ at dose $x_{ijk}$ (dose level $j$). It is the effect of agent $k$ in study $i$ at dose $x_{ijk}$ relative to placebo (or the minimum dose in the study $i$; see end of the section). If the log function instead of the logit is used to transform the probabilities, the model will estimate risk ratios instead of OR.

The parameter $\delta_{ijk}$, can be modelled then assuming a common- or exchangeable-effect model, see Table 2. For the common-effect model, the underlying true effect is assumed to be equal in all studies, so we set

$$\delta_{ijk} = \varDelta_{ijk}.$$

For exchangeable-effect model, $\delta_{ijk}$ are assumed to come from a common normal distribution with mean $\varDelta_{ijk}$ and variance $\tau^2$,

$$\delta_{ijk} \sim N(\varDelta_{ijk}, \tau^2)$$

The heterogeneity $\tau^2$ reflects between-studies variability, and it is assumed to be independent of the dose and agent. For multi-arm trials with more than one active agent examined, there are more than one $\delta_{ijk}$ per study, and as they are calculated using the same reference arm, they shall be jointly modelled using a multivariate normal distribution as in standard network meta-analysis [2].

To incorporate the dose-effect relationship in the model, we linked the parameters $\Delta_{ijk}$, to the transformed doses under an assigned function $F$, which we will call the dose-effect function:

$$\Delta_{ijk} = F\left(x_{ijk};\ \beta_{1,ik}, \beta_{2,ik}, \dots, \beta_{P,ik}\right) \tag{1}$$

The function $F$ can take various forms and the shape is defined by a set of $P$ parameters $\beta_{1,ik}, \dots, \beta_{P,ik}$. In addition to that, $F$ can be set differently for each agent $k$; $F_k$. Here we will set $F$ to be a RCS—the same for all agents.

The general form of the RCS with $\kappa$ knots $t_1, \dots, t_\kappa$ is defined as follows

$$F\left(x_{ijk};\ \beta_{1,ik}, \beta_{2,ik}, \dots, \beta_{P,ik}\right) = \tag{2}$$

$$\beta_{1,ik}\, x_{ijk} + \beta_{2,ik} f_2\left(x_{ijk}\right) + \dots + \beta_{(\kappa-1),ik} f_{(\kappa-1)}\left(x_{ijk}\right)$$

where for $m = 1, \dots, (\kappa - 2)$

$$f_{(m+1)}\left(x_{ijk}\right) = (x_{ijk} - t_m)_+^3 - \frac{t_\kappa - t_m}{t_\kappa - t_{\kappa-1}} (x_{ijk} - t_{\kappa-1})_+^3 + \frac{t_{\kappa-1} - t_m}{t_\kappa - t_{\kappa-1}} (x_{ijk} - t_\kappa)_+^3.$$

with $(x)_+ = x$ if $x > 0$ and 0 otherwise. For more details, see Section 2.4.5 in Harrell [18].

Setting three knots ($\kappa = 3$) will reduce $F$ in Equation 2 into a function with two coefficients. Then the dose-effect relationship becomes expressed by the linear and the spline terms; $\Delta_{ijk}$= $\beta_{1,ik} x_{iDk} + \beta_{2,ik} f\left(x_{ijk}\right)$. We use three knots for the remainder of the paper. A discussion about selecting the number of knots and their location can be found elsewhere [11, 18].

When the study $i$ does not have a placebo arm, we can choose an agent $R$ at the minimum dose level $r$ as the study-specific reference treatment. Then, the relative treatment effect $\Delta_{i(jr)(kR)}$ refers to the effect of agent $k$ at dose level $j$ versus agent $R$ at dose level $r$ ; it is modelled as

$$\Delta_{i(jr)(kR)} = F\left(x_{ijk};\ \beta_{1,ik}, \beta_{2,ik}, \dots, \beta_{P,ik}\right) - F\left(x_{irR};\ \beta_{1,iR}, \beta_{2,iR}, \dots, \beta_{P,iR}\right)$$

### *3.2.2 Dose-effect model across studies and agents*

To synthesise the dose-effect parameters $\beta_{p,ik}$ across studies, we employed the following assumptions (see Table 2). We can assume each agent-specific $p^{\text{th}}$ shape parameter $\beta_{p,ik}$ to be either exchangeable $\beta_{p,ik} \sim N(B_{p,k}, \sigma_{\beta,p}^2)$ (Assumption 2.1) or equal $\beta_{p,ik} = B_{p,k}$ (Assumption 2.2) across studies. We can simplify Assumption 2.1 by setting a common shape heterogeneity $\sigma_{\beta,p} = \sigma_\beta$.

Across agents, we can relate the shape parameters based on three possible assumptions (see Table 2). For agent $k$, we have a set of $P$ shape parameters $B_{p,k}$ ; it can be either independent $B_{p,k}$ (Assumption 3.1), it can have a common normal distribution $B_{p,k} \sim N(b_p, \sigma_{B,p}^2)$ (Assumption 3.2), or it can be fixed to a single value as $B_{p,k} = b_p$ (Assumption 3.3). The latter assumption requires a harmonisation of doses, so all agents' doses are measured on the same scale.

Let us define $\Delta_{.(ac)(AC)}$ as the expectation for the log-odds ratio between treatment $A$ at dose $x_a$ versus treatment $C$ at dose $x_c$. Now, to estimate the dose-effect curve between the two non-referent agents A and C, we can use consistency equations

$$\Delta_{.(ac)(AC)} = \Delta_{.aA} - \Delta_{.cC}$$
$$= B_{1,A}x_a + B_{2,A}f(x_a) - [B_{1,C}x_c + B_{2,C}f(x_c)] \qquad (3)$$

where $\Delta_{.aA}$ refers to the study-specific treatment effect of agent A at dose $x_a$ versus placebo.

### 3.3 Dose-effect network meta-regression model

We can extend the DE-NMA to DE-NMR by adding dose-covariate interaction terms. Assuming an RCS interaction between the covariate and the dose, the DE-NMA model (in Section 3.2) can be updated to

$$\text{logit}(p_{ijk}) = \begin{cases} u_i & \text{placebo} \\ u_i + \delta_{ijk} + Z_i F(x_{ijk};\ \gamma_{1,ik}, \gamma_{2,ik}, \dots, \gamma_{P,ik}) & \text{active agent} \end{cases}$$

The term $Z_i$ represents a study-level covariate. The parameters $\gamma_{1,ik}, \dots, \gamma_{P,ik}$ are expressing the impact of the dose-covariate interaction effect on the relative treatment effect. In most cases however, a linear interaction term should suffice (and would be estimable) so that $F(x_{ijk};\ \gamma_{1,ik}, \gamma_{2,ik}, \dots, \gamma_{P,ik}) = \gamma_{1,ik} x_{ijk}$.

Across studies, we can assume either exchangeable-effect; $\gamma_{m,ik} \sim N(\Gamma_{m,k}, \tau_\gamma^2)$ or a common-effect model; $\gamma_{m,ik} = \Gamma_{m,k}$. Across agents, we can model $\Gamma_{m,k}$ under one of the following three alternatives; estimate each one independently; $\Gamma_{m,k} = g_{m,k}$, assume exchangeable dose-covariate interaction terms $\Gamma_{m,k} \sim N(g_m, \tau_\Gamma^2)$, or presume a common dose-covariate interaction term $\Gamma_{m,k} = g_m$.

We assume consistency for dose-covariate interaction effects per treatment comparison, that is for the impact of dose-variable $m$ interaction on the parameter effect between two active agents $k_1$, $k_2$; it is $\Gamma_{m,\,k_1\,vs\,k_2} = \Gamma_{m,\,k_1} - \Gamma_{m,\,k_2}$. This means that when we assume $\Gamma_{m,k} = g_m$, $\Gamma_{m,\,k_1\,vs\,k_2} = 0$.

## 3.4 Dose-effect network meta-analysis model accounting for clusters

Often it might be desirable to group agents in classes and then estimate the class effect alongside agent effects. The assumptions for the shape parameters behind such a model are added as Assumption 3.4 and Assumption 3.5 in Table 2. Such parameters for agents $k_c$ belonging to class $c$ can be assumed either exchangeable $B_{p,k_c} \sim N\left(b_{p,c}, \sigma^2_{B,p}\right)$ or common $B_{p,k_c} = b_{p,c}$. Then the parameters $b_{p,c}$ are estimated independently for each class $c$.

When classes are considered, the doses of the agents within a given class need to be measured on the same (or equivalent scales) to calculate meaningful class-effects. For example, to estimate a dose-effect of all SSRIs, we will need first to transform the dose of each different SSRI into the same fluoxetine-equivalent scale.

## 3.5 Estimating an absolute mean effect for each agent at each dose level and calculating a treatment hierarchy.

With many treatments and doses, results are more easily presented and understood using absolute estimands, such as the response probability $P_{jk}$ for a specific dose $j$ of a certain agent $k$. In a Bayesian setting, this can be done by combining the estimated dose-effect parameters with the response probability for placebo $P_0$. The latter can be computed outside the DE-NMA model by placing a binomial distribution for the corresponding events $r_{i0}$ with sample size $n_{i0}$ and probability of the event to occur in placebo arm $p_{i0}$

$$r_{i0} \sim Binom(p_{i0}, n_{i0}),$$

$$logit(p_{i0}) \sim N(logit(P_0), \sigma_0^2).$$

Next the predicted probability of the event to occur at dose $j$ and agent $k$ is

$$P^*_{jk} = \text{expit}\Big(logit(\tilde{P}_0) + F\left((x_{jk};\, \tilde{B}_{1,k}), (x_{jk};\, \tilde{B}_{2,k}), \ldots, (x_{jk};\, \tilde{B}_{p,k})\right) + \tilde{g} \times cov_{pred} \times x_{jk}\Big),$$

where the tilde (~) refers to the posterior of parameters.

These probabilities may be used then to rank the agents according to their efficacy at each dose level. However, to make comparison easy, one might need to transform the doses into a single scale using equivalence formulae, if available.

# 4 Application in dose-effect of antidepressants

## 4.1 Implementation of the models and diagnostics

We conducted a DE-NMA under five different model specifications. M1 is the primary dose-effect NMA model, and then we added three dose-effect NMR models (M2 to M4) for covariates, risk of bias (low versus high), study publication year (centred at 2010), and the variance of logOR (to evaluate small study effects). In M5, we accounted for class effects instead of the agent effects as listed in Appendix Table 1. All models employ assumptions 1.1 and 2.2. (Table 2). M1-M4 additionally employ assumption 3.1. We set a common dose-covariate interaction effect across studies and agents in M2-M4 ($\Gamma_{m,k} = g_m$). In M5 class effects are modelled using assumption 3.5 where all doses are transformed to fluoxetine-equivalent dose using previously established transformation.

We modelled the dose-effect relationship with RCS with three knots. Because agents have different dose ranges, knots are placed for each agent at 25%, 50%, and 75% percentiles of the corresponding observed dose range. We investigated the sensitivity of the estimated curve to knots position, only for M1 by placing knots at 10%, 20%, and 30% percentiles.

All parameters were estimated using JAGS program which is implemented via R. We assessed the overall performance of the model using the Deviance Information Criterion (DIC) statistic and leverage plots. The values of DIC can be used to compare between different models but they need to have the same likelihood and data. The model provides the best balance between model fit and complexity when it has the lowest DIC.

We estimated the parameters with Markov Chain Monte Carlo (MCMC) using three chains with $1 \times 10^4$ iterations, $4 \times 10^3$ burn-in, and a thinning of one. We set a minimally informative prior for the placebo effect $u_i \sim N(0, 10^3)$ and the shape parameters $B_{p,k} \sim N(0,10^3)$, $b_p \sim N(0,10^3)$. The two heterogeneity parameters are given a uniform prior $\tau, \sigma_{p,B} \sim$ Unif(0,5). For the covariate effect in DE-NMR (M2-M4), we set $g_m \sim N(0,10^3)$. For the placebo response model, we placed $\text{logit}(P_0) \sim N(0, 10^3)$ and $\sigma_0 \sim \text{Unif}(0,5)$.

We used the `rcs` function from the `rms` package to compute the RCS transformations [19]. The codes are available at Zenodo library [15]. We used different numerical and graphical methods (using the `coda` package [20]) to investigate the convergence of the MCMC. The results are provided as a posterior median with the 95% credible interval (CrI).

## 4.2 Results

In Figure 2, we depict the absolute dose-effect relationship for each antidepressant along with the overall placebo effect for M1. The response to placebo is estimated at mean 36.2% (95% CrI $34.4\% - 38.0\%$) (blue line). All antidepressants are more effective than placebo after some dose level which differed by agent. However, for some agents, there is a lot of uncertainty particularly at high dose levels (except clomipramine at a low dose level where we have no data). The efficacy initially increased up to a specific dose only to flatten out after a given dose for most agents. For example, the efficacy of duloxetine increased until 75 mg, then it leveled out after that. We identified moderate to small differences in the estimated curves from M1 when we changed knot positions (see Appendix Figure 6). However, the overall conclusions do not change with the change in knot locations.

In M2 to M4 models, we estimate the dose-effect curves assuming three different covariates. In Appendix Figure 3, M2 suggests that studies with high risk of bias (RoB) tend to overestimate on average the efficacy compared to low RoB for some drugs such as bupropion. The average of bupropion efficacy is also more exaggerated in older studies (M3); this is additionally observed for some other antidepressants, see Appendix Figure 2. In M4, the efficacy of most antidepressants is on average higher in small studies (or studies with large variance of logOR) compared to studies with large sample size (Figure 3). In Appendix Table 5 we summarize the findings and the performance for all models.

In Appendix Figure 5, we present the contribution of each observation to $p_D$ in y-axis and to $\overline{D}_{res}$ in x-axis along with the overall model fit measures DIC, $p_D$ and $\overline{D}_{res}$. The DIC is 790 for the M1, M2, M3 models and it is slightly declined to 789 for M4 model with the variance of logOR as a covariate (Appendix Table 5 ).

In Figure 4, we show the absolute probabilities under the class effect model M5. As expected, for classes with many drugs such as SSRIs and serotonin-norepinephrine reuptake inhibitors (SNRIs) we gained precision compared to the agent-level models M1.

# 5 Discussion

We present a dose-effect NMA model to synthesize evidence from trials that compare multiple agents at different dosages. To model the dose-effect relationship, we choose RCS to take advantage of their flexibility. We added two extensions to the model: a dose-effect network meta-regression to account for study-level covariates and groups of agents in a class-effect model. We implemented various DE-NMA models in a network of antidepressants and placebo, and the resulting dose-effect shapes are in line with clinical expectations and previous findings [4, 11, 17]. Introducing covariates allows us investigating how the dose-effect curve changes at different values of the covariate. These changes are substantial for antidepressants when we added the logOR variance as a covariate. Modelling class effects resulted in more precise estimates of the dose-effect association. We, additionally, identified the specific dose range in which antidepressant effect exceeds the placebo effect and beyond which dose the effect no longer increases.

Some limitations of the DE-NMA models need to be acknowledged. First, the findings from such analyses can be sensitive to the assumptions about the dose-effect shapes (whether it is an assumed polynomial or splines). Besides a sensitivity analysis, researchers can *a priori* narrow down the set of possible shapes to the ones that best reflect the known biological behaviour of agents. If needed, the goodness of fit statistics can guide the final choice when enough data is available. When several models provide equally good fit, Bayesian model averaging can be used. The location of knots in RCS requires particular attention. The estimation of the model could be sensitive to the location of knots; and a sensitivity analysis is recommended to explore any impact on the results [11]. Although some researchers argue that the location of knots is not problematic in general [18, 21], we have previously found that positioning the knots at places where shift changes in the effect are expected might be a good strategy [11].

Second, there are often very few observations for the same agent to estimate the dose-effect relationship with precision high enough to inform clinical practice. In such cases, the analysis might require considering other sources of information, such as informative priors for the shape, and coefficients of the association based on an external source, such as observational studies, or clinical expertise. Alternatively, we can impose stronger assumptions by borrowing information internally, such as assuming class-effects or even exchangeable dose-effect coefficients across all agents. This assumption will improve the parameters' identifiability, and it also enables us to analyse a disconnected network. This approach, however, requires the doses to be harmonized across the agents and assumes exchangeable dose-effect shapes across agents, which might be difficult to justify in practice.

At present, we only synthesized fixed-dose studies. Studies with a flexible dose schedule, where the dose is increased up to a maximum targeted level, depending on the patients' response and acceptability, require special attention. The analysis of post-randomisation dose adjustments requires causal modelling and individual participant data. Synthesizing fixed and flexible-dose studies is challenging and results will require careful interpretation. Finally, we did not examine potential inconsistencies in the data; this can be done using newly introduced methods [22].

Technically, dose-response meta-analysis with RCS with three knots require two studies with at least one of them having three different dose levels. However, issues of precision, model fit, and heterogeneity question the utility of results from such analyses. Depending on the sparseness of the outcome and the complexity of the underlying dose-response shape, a substantial amount of data might be required to obtain useful results from dose-response meta-analyses.

In the present study, we only synthesized fixed-dose randomized studies where all patients in a study arm were prescribed and took the same dose of the same antidepressant. That means that dose is not a "patient-level" characteristic aggregated over the study arm, but an arm-level characteristic. Consequently, aggregation bias is unlikely in the dose-response association with fixed-dose randomised trials. Including studies with a flexible dose schedule, where the dose is increased up to a maximum targeted level according to the patients' response and acceptability, warrants special attention. The analysis of post-randomisation adjustments of the dose requires causal modelling and individual participant data. Synthesizing fixed and flexible-dose studies is challenging and results need cautious interpretation.

There is a variety of functional forms to model the dose-effect relationship in NMA, such as the Emax model [8]. The Emax model is widely used in drug development context where the focus is on studying drug safety and finding optimal doses (e.g., finding the dose at which half of the maximum effect is achieved). In clinical practice, however, the interest is on estimating the dose-effect relationship for the whole dose range. In this context, the parameters of Emax model are of less interest and the dynamics of the function makes it less likely to portray the underlying true dose-effect association. In contrast, the RCS offer sufficient flexibility to capture the biological behaviour of agents with only few parameters (only two parameters when we set three knots). This is particularly important in larger dose levels where the efficacy of many pharmacological agents is expected to level out.

Fractional polynomial is another alternative to model the dose-effect relationship. They have been shown to perform well when modelling longitudinal data in NMA [23] but have not been implemented in DE-NMA context yet. However, they can be less appealing when modelling dose-effect associations. Fractional polynomials are non-local functions which means they can be less efficient in detecting the multiple changes in drug dynamics [24].

Although fractional polynomials might be useful in the dose-findings studies where focus is on the safest dose [25], the RCS might be preferable in (network) meta-analysis contexts where we can benefit from the locality of the RCS to place knots at the expected changing points based on clinical or biological knowledge.

Little work has been done systematically comparing the performance of various functions in the dose-response context. Zhang et al. [26] conducted a dose-effect meta-analysis to model the sleep duration and the risk of all-cause mortality, assuming different dose-effect shapes. They found that RCS performed well, while fractional polynomials yielded unreasonable results at five and six hours of sleep. Additionally, fractional polynomials need intensive computations to find the optimal powers, which is cumbersome to implement in a Bayesian setting. Further work is needed in this direction to study and compare different dose-effect shapes and pinpoint the advantages and limitations of the fractional polynomials in this context.

Our study's model is an extension of our previous work in pairwise meta-analysis [11]. Dose-effect pairwise meta-analysis models require transforming the doses into a common scale across agents, which is not always straightforward or even possible. DE-NMA allows us to compare multiple agents simultaneously, using their original doses. It can also answer key questions about what treatments are preferable and what dose can maximise the relative effects. These results from the DE-NMA model are important for drug guideline developers, health technology assessment agencies, and of course patients and their treating clinicians.

*Table 1* Notations for dose-effect network meta-analysis (DE-NMA).

| | |
|---|---|
| $i = 1, \dots, ns$ | Study id |
| $j$ | Index for the dose levels in study $i$ |
| $k = 1, \dots, K$ | Agent |
| $c = 1, \dots, C$ | Exposure clusters |
| $p = 1, \dots P$ | Number of dose transformations associated with the dose-response shape. For a linear shape $p = 1$ and for quadratic and restricted cubic splines $p = 2$ |
| $x_{ijk}$ | The $j$-th dose in study $i$ for agent $k$ |
| $x_{dk}$ | Minimal dose $d$ for agent $k$ |
| $Z_i$ | covariate in study $i$ |
| $r_{ijk}$ | Number of events in dose $j$ within study $i$ for agent $k$ |
| $n_{ij}$ | Sample size in dose $j$ within study $i$ for agent $k$ |

*Table 2 List of potential assumptions for the parameters in DE-NMA model.*

| |
|---|
| Assumptions about the effect parameter $\delta_{ijk}$ |
| Assumption 1.1 - exchangeable <br> $\delta_{ijk} \sim N(\Delta_{ijk}, \tau^2)$ |
| Assumption 1.2 - common <br> $\delta_{ijk} = \Delta_{ijk}$ |
| Assumptions about the p$^{th}$ within-study shape parameter $\beta_{p,ik}$ |
| Assumption 2.1- exchangeable <br> $\beta_{p,ik} \sim N(B_{p,k}, \sigma^2_{\beta,p})$ |
| Assumption 2.2 – common <br> $\beta_{p,ik} = B_{p,k}$ |
| Assumptions about the p$^{th}$ between-agents shape parameter: $B_{p,k}$ |
| Assumption 3.1 – independent <br> $B_{p,k} = b_{p,k}$ |
| Assumption 3.2 - exchangeable <br> $B_{p,k} \sim N(b_p, \sigma^2_{B,p})$ |
| Assumption 3.3 - common <br> $B_{p,k} = b_p$ |
| Assumption 3.4 – exchangeable class-effect across agents $k_c$belonging to class $c$ $B_{p,} \sim N(b_{p,c}, \sigma^2_{B,p})$ |
| Assumption 3.5 – common class-effect across agents $k_c$belonging to class $c$ <br> $B_{p,k_c} = b_{p,c}$ |

# Figures

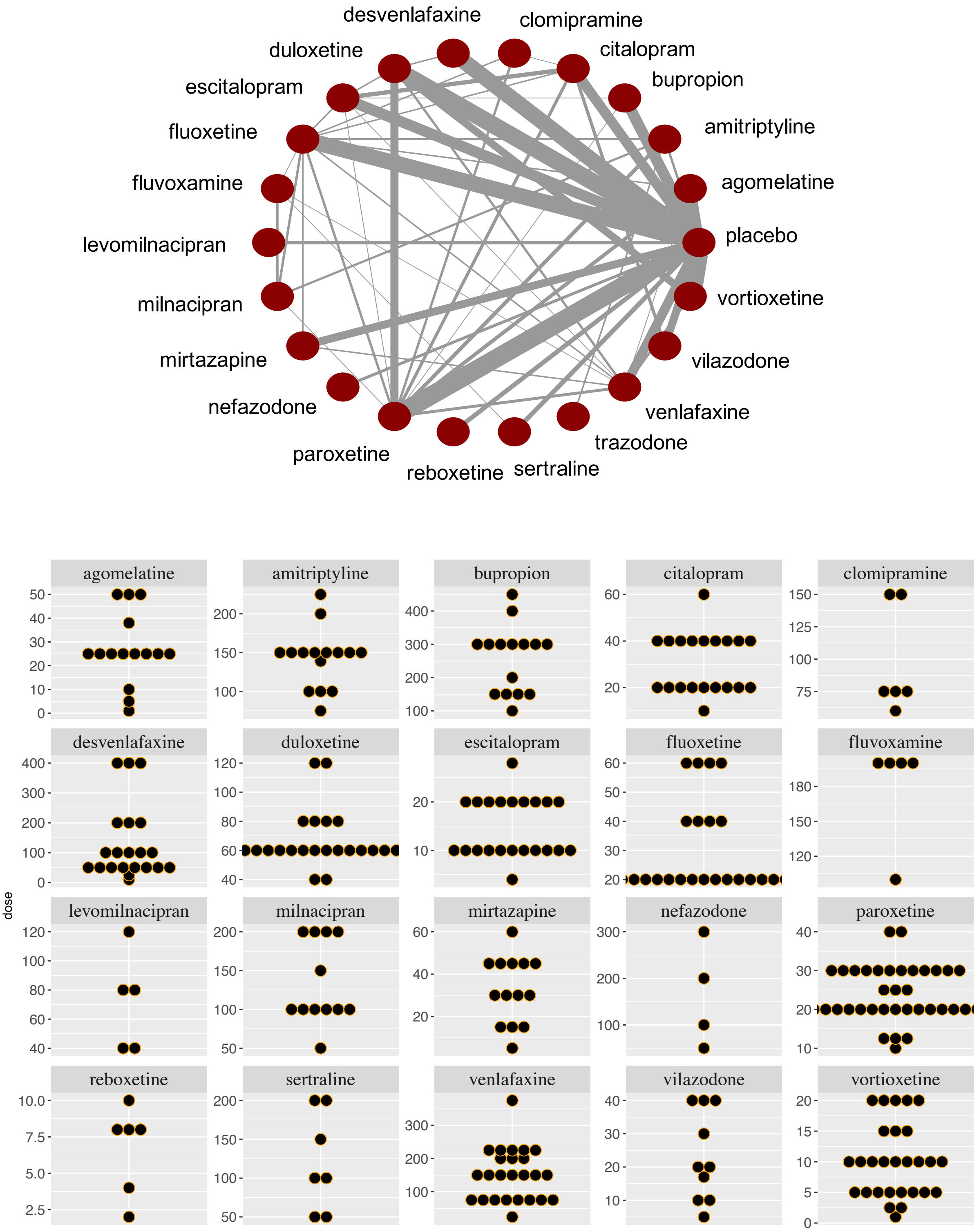

*Figure 1 (a) Network meta-analysis of studies comparing 21 antidepressants and placebo. The width of the lines is proportional to the number of trials comparing each pair of agents. This plot was produced using the plot() function from the R package MBNMAdose (b) The dose distribution for the 21 antidepressants.*

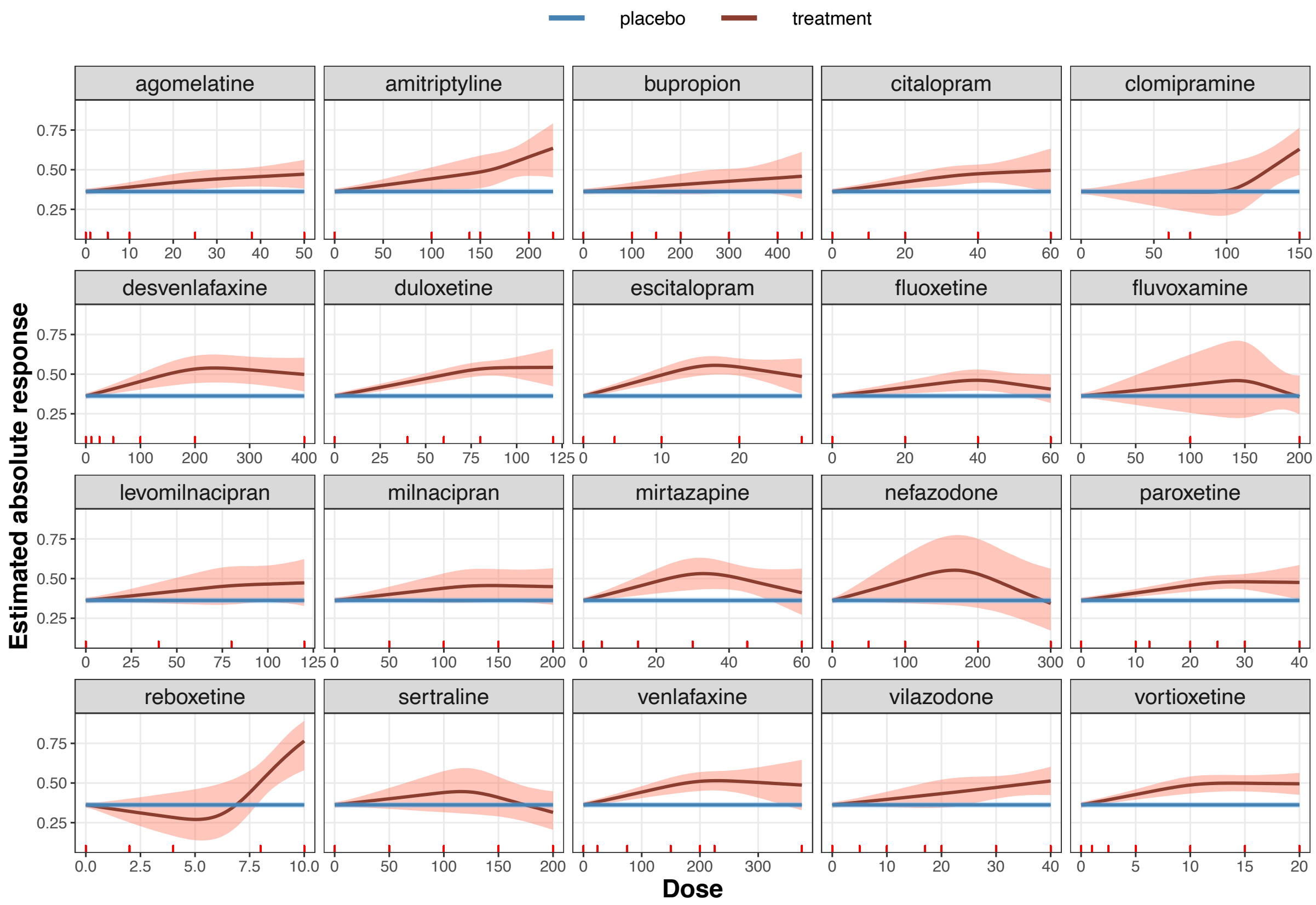

*Figure 2 Dose-effect network meta-analysis summary curve for each antidepressant. The blue line depicts the effect estimated from all placebo arms in the network (36.2%) and its 95% credible region. The red line represents the absolute response to each antidepressant (estimated from model M1) and the shaded area is its 95% credible region.*

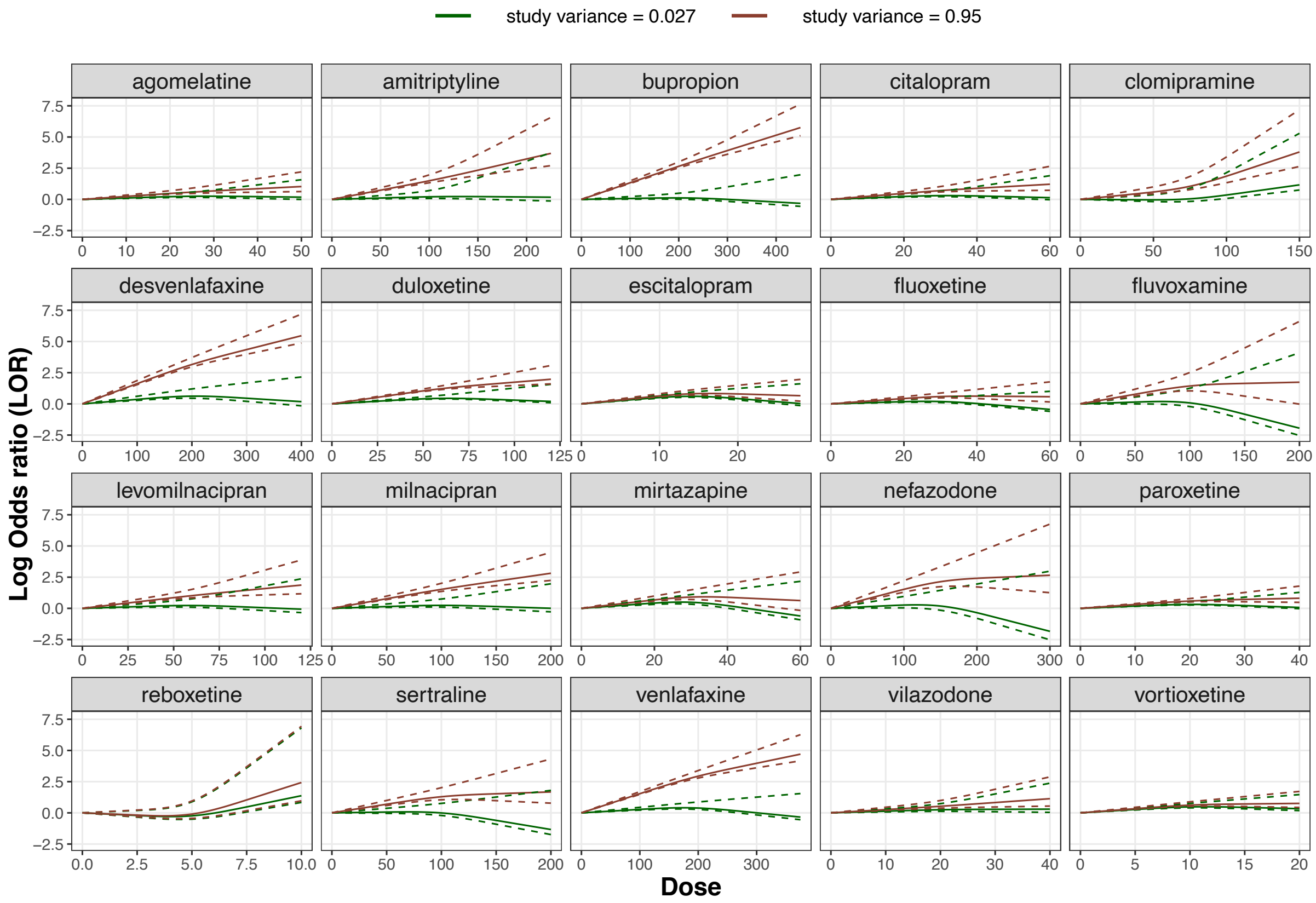

*Figure 3 Dose-effect network meta-regression summary curve of each one of the 20 antidepressants using the study variance of log odds ratio as a covariate (estimated from model M4). The dose-effect curves are depicted for studies with low variance at 0.027 (green) and with large variance at 0.95 (red). The dotted lines represent the 95% credible interval.*

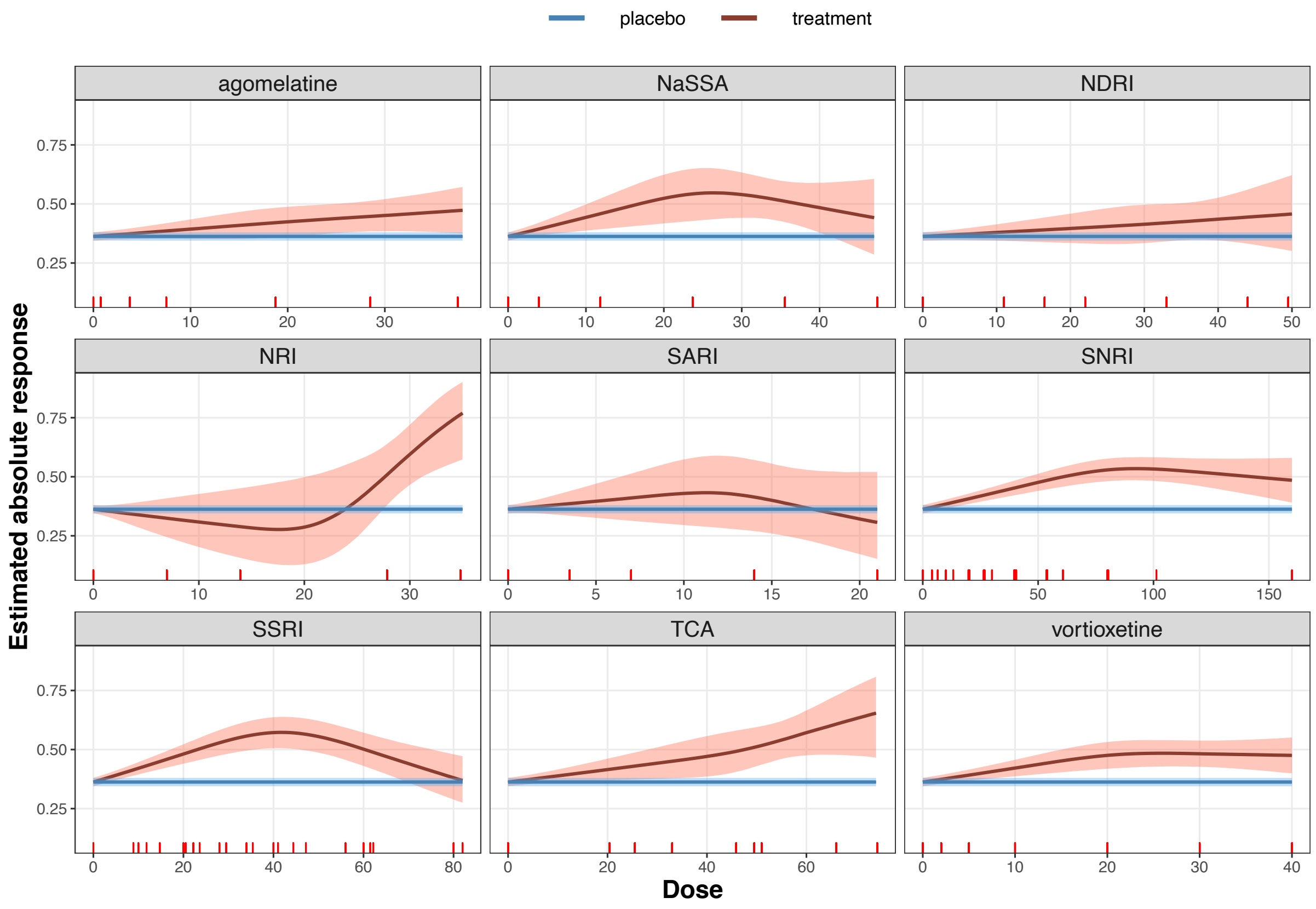

*Figure 4 Dose-effect network meta-analysis summary curve for each of the 9 drug classes (see Appendix Table 1). The blue line depicts the effect estimated from all placebo arms in the network (36.2%) and its 95% credible region. The red line represents the absolute response to each drug class (estimated from model M5) and the shaded area is its 95% credible region.*